\documentclass[aps, preprint, showpacs, superscriptaddress]{revtex4}

\usepackage{graphicx}
\usepackage{amsmath}
\usepackage{bm}
\newcommand{\be}{\begin{equation}}
\newcommand{\ee}{\end{equation}}

\begin{document}
\title
%---------------------------------------------------------
{Adiabatic and Nonadiabatic Electronics of Materials}
%---------------------------------------------------------

\author{V.A. Vdovenkov}

\affiliation{Moscow State Institute of Radioengineering,
Electronics and Automation (technical university)\\
Vernadsky~ave. 78,~117454~Moscow,~Russia}

\begin{abstract}
Physical foundations of adiabatic and nonadiabatic electronics of materials are considered in this article. It is shown the limitation of adiabatic approach to electronics of materials. It is shown that nonadiabatic physical properties of solid materials (hyperconductivity, superconductivity, thermal superconductivity and phonon drag of electrons at Debye's temperatures of phonons) depend on oscillations of atomic nuclei in atoms.  
\end{abstract}

\pacs{63.20.Kr, 63.90.+t, 71.00.00, 72.20.Pa}
%71.00.00   Electronic structureof bulk materials
%63.20.Kr   Phonon-electron and phonon-phonon interactions
%63.90.+t   Other topics in lattice dynamics
%72.20.Pa   Thermoelectric and thermomagnetic effects
%74.20.Mn   Nonconventional mechanisms (Superconductivity)(spin fluctuations, polarons, ets.) 

\maketitle

\section{Introduction}
In modern electronics of materials, adiabatic principle mainly is used. According to this principle, exchanging of energy between electrons and nuclei of atoms are neglected. Today's dominating electronics in essence is adiabatic electronics. It allows to consider materials approximately, in adiabatic approach, and to study limited circle of their physical properties. Development of nonadiabatic electronics, on the contrary, just begins. It takes into the account processes of energy exchange by between electrons and nuclei of atoms, has no basic restrictions, can reduce or remove problems of modern electronics and reveal new physical properties of materials.

This article is devoted to analysis of physical foundations of adiabatic and nonadiabatic electronics of materials and properties of boson-fermion fluid, arising under nonadiabatic conditions in materials.
\section{Adiabatic electronics of materials}
Bases of adiabatic electronics were stated by M. Born and R. Oppenheimer \cite{Born1927} in the solution of Sñhrodinger's stationary equation for a material $H\Psi=W\Psi$, where Hamiltonian $H$ contains operators of kinetic energies of electrons $T_e$ and atomic nuclei $T_z$ , and also the crystal potential V is dependent on sets of electrons coordinates (r) and atomic nuclei coordinates (R) in a material. According to M. Born and R. Oppenheimer this equation is equivalent to the following two equations:
\begin{equation}
\label{eq1}
(T_e + V)\phi(r,R) = E\phi(r,R),
\end{equation}
\begin{equation}
\label{eq2}
(T_z + E + A)\Phi(R),
\end{equation}
where $T_e$  - kinetic energy of electrons' system, $W$ - energy of a material,  $\phi(r,R)$ and  $\Phi(R)$ -  wave functions for electrons system  and  nuclei system, 
\begin{equation}
\label{eq3}
A=-\sum_I(\hbar^2/M_I)\int\phi^*(r,R)\nabla^2_R\phi(r,R)d\tau
\end{equation}
- adiabatic potential, $M_I$ - mass of nucleus in $I-th$ atom, $d\tau$  - an element of the material's volume. Potential $A$ describe exchange of energy between electrons and nuclei. In adiabatic approach, potential $A$ is small comparing to $W$, and it is neglected. One believe that if $A\equiv 0$  then the exchange of energy between electrons and nuclei is impossible, then, one believes that the problem of a material, stated by the equations (\ref{eq1}, \ref{eq2}), is an adiabatic problem, and electronic properties, described by the equation (\ref{eq1}), represent a theoretical basis of adiabatic electronics of materials. However, one can see from equations (\ref{eq1}, \ref{eq2}) that upon exception of $A$, a full separation of variables in these equations  is not achieved, as  $r$ and $R$  are arguments of $V$. Hence, the exchange of energy between systems of electrons and nuclei in adiabatic approach, generally speaking, is not excluded.  

So, physical parameters of crystalline silicon were calculated in adiabatic approach by G. Pastore, E. Smargiassi and F. Buda \cite{Pas1991}. It turned out, that kinetic and potential energies of electrons system and nuclei system deviate from average values in opposite phases to each other with characteristic frequencies. It means that the energy exchange between electrons and nuclei system occurs even at adiabatic conditions. Autors \cite{Pas1991} supposed that the greatest among the frequencies was the result of numerical method of calculation, because at that time such frequency oscillations in crystals were not known. Later it was found out, that this frequency coincides with the sum of two frequencies (frequency of inherent oscillations of nucleus in silicon atom minus frequency of crystal phonon). In other words, inherent oscillations of atomic nuclei are the fundamental property of manerials.  Amplitudes of nucleus oscillations are close to $10^{-12}$ meter~ and consequently processes connected to them refer to \textit{subangstroem} electronics of materials.  These oscillations in particular can be an energy source for chemical reactions and for phase transitions even at extremely low temperatures. The given result allows us to consider the adiabatic principle differently. Adiabatic condition ($A\equiv0$) means impossibility of the systematical, unidirectional stream of energy from electrons to nuclei or from nuclei to electrons only, but does not exclude the stream of energy periodically changing its direction. 

P. Dirac \cite{Dirac1930} investigated the time-dependent Schrodinger's equation for a crystal and studied an opportunity of co-ordinates separation for electrons and nuclei that correspond to adiabatic principle. He has shown that electrons and atomic nuclei move within the effective self-consisted potential fields and generally their movements cannot be described by equations, independent from each other. Therefore, the adiabatic principle in materials is approximate, permitting to use it with some accuracy in certain conditions, and its applying each time requires substantiation and estimation of possible mistakes arising during calculation of physical quantity in such an approach.  

The regular error of energy calculation for electrons in adiabatic approximation is about $<A> \approx 10^{-5} W$. It turns out to be comparable with widths of forbidden energy gapes in many semiconductors and therefore is not always allowable. In particular, the given error can be one of the reasons of inexact definition of deep energy levels of the local centers in semiconductors, calculated under adiabatic approximation.
  
The amendments to energy of the crystal, that arise as a result of removal of potential eq. \eqref{eq3} out of equation \eqref{eq2}, according to M. Born \cite{Born1927}, are proportional to the integer powers of small parameter $\eta=(m/M_I)^{1/4}<<1$ , where  $m$- electron mass. It has given an occasion to justify any applying of adiabatic approximation, based on smallness of $\eta$, proved to be wrong, and, as a matter of fact, it is generally wrong in all cases. Nonadiabatic principle and terms of its applying to materials have other physical sense. 
\section{Conditions for applying adiabatic approach}
It was shown by C. Herring \cite{Herr1956} (see else: \cite{Stone1975}), that adiabatic approach can be applied, if
\begin{equation}
\label{eq4}
E_{kl} >> \sum_\mu \hbar  \omega_\mu~\Delta R_\mu \int d^3 r ~ \phi^*_k\frac{\partial}{\partial R_\mu}~ \phi_l~~,
\end{equation}
where $E_{kl}$ - energy of the allowed electronic transitions between states  $k$ and $l$ ,  $\Delta R_\mu$ - characteristic displacement of nuclei system on frequency $\omega_\mu$ , $\mu$  - type of oscillations, $\phi_k$ and $\phi_l$ - electron wave functions, $\hbar$ - Dirac's constant. Adiabatic approach was studied by A. Davidov \cite{Dav1973} in conditions when oscillations of atomic nuclei are allowed only with one frequency  $\omega_\mu$. He has come to a conclusion that adiabatic approach can be used if 
\begin{equation}
\label{eq5}
E_{kl}>>\hbar\omega_\mu. 
\end{equation}
In other words, the adiabatic principle can be soundly applied if energy of nuclei oscillations is less than energy of the allowed electronic transitions, for example if the energy of nuclei oscillations is less than width of the forbidden energy gape of a semiconductor. The condition eq. \eqref{eq5} is sufficient, but is not a requirement. Therefore in some cases applying of adiabatic approximation is justified though condition eq. \eqref{eq5} is not carried out.

Thus, criterion of applicability of adiabatic approach in materials is not so simple, as it quite often looks. The adiabatic principle dominates in modern physics of materials, but in some cases it is used unreasonably. Implications of adiabatic theories quite often appear unproductive because of inaccurate applying of adiabatic approach. On the contrary, using of nonadiabatic principle allows to reveal and to use new properties of materials. So, conditions for applicability of adiabatic approach stated by C. Herring \cite{Herr1956} and A. Davidov \cite{Dav1973} contain various types and frequencies of nuclei oscillations. Meanwhile, these nuclei oscillations (unlike oscillations of atoms or ions) are poorly investigated. It hinders well-founded applications of adiabatic approach and halts development of nonadiabatic electronics.

\section{Types and frequencies of inherent oscillations in atoms of materials}
\begin{figure}[ht]
\vspace*{0cm}
%\begin{center}
\includegraphics[width=8cm]{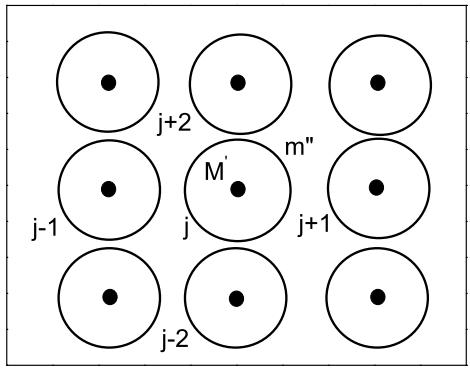}
%\end{center}
\vspace*{0cm}
\caption{Crossection of a three-dimensional crystal model by a plane.}\label{fig1}
\end{figure}

For determining types and frequencies of nuclei inherent oscillations in atoms, it is expedient to use model of a crystal in which each atom is represented by an electron shell and a nucleus connected with each other by quasi-elastic force.  Crossection of such a three-dimensional model by a plane containing centers of some electronic shells is shown on Fig.~\ref{fig1}.
Electron shells with masses $m"$ on Fig.~\ref{fig1} are represented as circles, nuclei with masses $M'$ are represented by dark circles in centers of electronic shells. According to analytical mechanics the frequency of normal oscillations of a nucleus in $j-th$ atom can be defined if to suppose absolutely rigid all quasi-elastic connections in considered model except one quasi-elastic connection acting between a nucleus and shell in a $j-th$ atom. In such conditions the oscillations of a crystal are represented by oscillations of a nucleus relatively motionless electronic shell of $j-th$ atom.  Physically equivalent situation is established out if mass of an electronic shell of $j-th$ atom $m" =\infty$.  Hence, the frequencies spectrum of nucleus oscillations is defined by known mass of a nucleus and by the potential (electric) field which hold a nucleus in neighbourhood of the center of electron shell. This electric field is formed by coulomb fields of nuclei and electronic shells of all atoms of a crystal. In our case it is enough to take into account a field created only by shell of $j-th$ atom in a neighborhood of her center. Thus, frequencies of nucleus oscillations can be defined by considering a movement of a nucleus in a field of a motionless environment that is in adiabatic conditions.

One can see from equation \eqref{eq2}, that in adiabatic approach ($A\equiv0$) and the potential field acting on a nucleus, coincides with full electrons energy which for neutral atom can be written down so:
\begin{equation}
\label{eq6}
E=T_e + E_{Ze} + E_{ee} + E_{ex}~~,
\end{equation} 
were $T_e$  -  kinetic energy of electrons, $E_{Ze}$  - energy of electrons attraction to a nucleus,  $E_{ee}=\frac{1}{2}\int \Phi (r) \chi (r) d \tau$  - energy of their  mutual pushing away from one enother, $E_{ex}$  - exchange energy. Electronic density $\chi(r)=e \sum^Z_{i=1}|\phi_i(r)|^2$, $\Phi(r)$ - electrostatic potential, created by electron shell in a point $r$, $d \tau $ - differencial of material volume where integration is carried out, $e$ - electron carge, $Z$ - atomic number, $i$ - elecron number. Cyclic frequency of harmonious oscillations of a nucleus in field, described by \eqref{eq6} is equal $\omega=(\beta/M)^{1/2}$, where $\beta$ - factor of the qusai-elastic force connecting a nucleus with his environment. This frequency $(\omega)$, as is well known, is equal to a difference of the neighbor oscillations frequencies in spectrum of quantum harmonious oscillator.  We take it into account and begin the frequencies calculation of nuclei inherent oscillations in various atoms.

In the atom of hydrogen (with atomic number $Z = 1$) $E_{ee}=0$, $E_{ex}=0$,  $T_e=-E_{Ze}/2$ under the virial theorem and $E=E_Ze/2$. The normalized wave function in the basic state of hydrogen atom is $\Psi_{1s}=(1/\pi a^3)^{1/2}exp(-r/a)$, where  $a$ - Bohr's radius. By integrating twice the Poisson equation $\frac{1}{r} \frac{d^2}{dr^2} (r\Phi)= \frac{-e}{\epsilon_0} |\Psi_1s|^2$ at boundary conditions $\Phi(r=\infty)=0$ and $\Phi(r=0)=\textit{const}$ we receive~ $\Phi(r)=(e/(4 \pi \epsilon_0 a^3))[(1/a+1/r)exp(-2r/a)-1/r]$ and $E=(e\Phi(r))/2$, where $\epsilon_0$ - the electric constant. We expand $E$ in power series, reject all terms containing $r$ in degrees are higher than two, and determine parabolic potential $E^"(r)$ in which oscillations of a nucleus are harmonious. Then we calculate $\beta=(d^2E^"(r)/dr^2)_{r=0}=e^2/(6 \pi \epsilon_0 a^3)$. Further we calculate elementary quantum of nucleus oscillations in the atom hydrogen~ $\hbar\omega_1=\hbar \sqrt{\beta/m_p} \cong 0.519 eV$, where $m_p$ is mass of proton. 

 Energy of three-dimensional quantum oscillations of a nucleus are described by the formula
\begin{equation}
\label{eq7}
E(\nu)=\hbar \omega_Z [(1/2+\nu_1)+(1/2+\nu_2)+(1/2+\nu_3)]~~,
\end{equation}
where $\nu_1, \nu_2, \nu_3$ - the oscillatory quantum numbers independently accepting values 0, 1, 2,...

In helium atom  $(Z=2)$ two electrons in the basic state have wave functions $\Psi=\phi_{1s}(1)\phi_{1s}(2)=\frac{1}{\pi}(Z^*/a)^3 exp[(-Z^*/a) (r_1+r_2)]$, where $Z^*=2-5/16\cong1.6875$ - the effective charge of the nucleus not equal to 2 owing to shielding of the nucleus by electrons (see ref.: \cite{Dav1973}, p. 338). We simplify the equation \eqref{eq6} using the property of two-electronic systems for which $E_{ex}=-E_{ee}/2$. $E=T_e +(Ze/4)\Phi(r)$. Under the virial theorem $E=(Ze/8)\Phi(r)$. Similarly to calculations for the hydrogen atom, we integrate Poisson equation with electronic density $e|\Psi|^2$,  determine $\Phi(r)$, and determine $\beta=(Ze^2)(24\pi\epsilon_0)^{-1}(Z^*/a)^3$. Then we determine quantum of inherent oscillations of nucleus in helium atom~  $\hbar\omega_2=\sqrt{(Ze^2)(24 \pi \epsilon_0)^{-1}(Z^*/a)^3[2(m_n+m_p)]^{-1}}\cong 0.402 eV$, where $m_p$ and $m_n$ - masses of a proton and a neutron.

Potential field in many-electron atoms is spherically symmetric and the normalized radial wave function ($R_{nl}$) of any electronic state can be expressed through hypergeometrical function F(b,c,d) (see reference: \cite{Dav1973}, p. 179):
\begin{equation}
\label{eq8}
R_{nl}=N_{nl}(\frac{2Zx}{n})^lF(l-n+1,2l+2,\frac{2Zx}{n})exp(\frac{-Zx}{n})~,
\end{equation}
where $N_{nl}=\frac{1}{(2l+1)!}\sqrt{\frac{(n+1)!}{2n(n-l-1)!}(\frac{2Z}{n})^{3/2}}$, $x=\frac{r}{a}$, $n$ - the principal quantum number, $l$ - the orbital quantum number. It follows from the \eqref{eq8}, that electronic density about a point of shell center is created mainly by $s-$ electrons but densities of $p-, d-, f-, ...$ electrons are insignificant. The density of $s-$ electrons from L, M, N states $( n = 2, 3, 4...)$ is supplemented to density of K - electrons $(n = 1)$. The share of density from these conditions can be defined as squares of radial wave functions ratios: $R_{20}/R_{10}^2\cong 0.125$; $(R_{30}/R_{10})^2\cong0.037$; $(R_{40}/R_{10})^2\cong0.0123$. Thus, it is visible, that the contribution to electronic density created by $2s-$, $3s-$, $4s-$ electrons give approximately 17.4 pecent that can cause increase in frequencies of nucleus oscillations about 5 percent. In the many-electron atoms one take into account screening of a nucleus charge by electrons, by using an effective charge of a nucleus $Z^{**}=Z-\mu$, where $\mu=\sigma Z^{1/3}$ and values $\sigma$ differ from unit a little for different atoms: \cite{Flug1971}, vol. 2, p. 153. In view of these data the energy quantum of nucleus inherent oscillations (quantum $\alpha$~- type of inherent oscillations) in the atom with number $Z > 2$ is
\begin{equation}
\label{eq9}
\hbar\omega_Z=\hbar\omega_2\sqrt{{(Z-5/16-\mu)}^3\Lambda(Z-\mu)/Z}~,
\end{equation}
where~ $\hbar\omega_2=0.402 eV$ - quantum of inherent oscillations of a nucleus in helium atom, $\Lambda=1.2$ takes into count influence of electronic $s-$ states with quantum numbers $n > 1$ on value~ $\hbar\omega_Z$ at $\alpha$~- type of inherent nuclei oscillations. The same result appears, when the theorem about ellipsoid's potentials is applied to the electron shell \cite{Mur1976}, according to the theorem inside regular intervals the potential of a charged ellipsoid is constant. Energy spectrum of nucleus inherent oscillation in the atom with number $Z$ is described by the formula \eqref{eq7} where~ $\hbar\omega_Z$ is calculated under the formula \eqref{eq9}.

\begin{figure}[ht]
\vspace*{0cm}
%\begin{center}
\includegraphics[width=8cm]{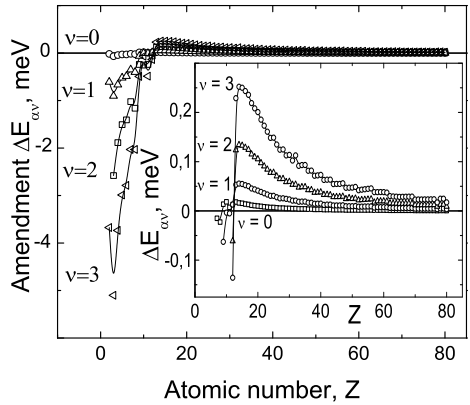}
%\end{center}
\vspace*{0cm}
\caption{Amendments $(\Delta E_{\alpha\nu})$ to energy of $\alpha-$ type oscillations in conditions with  $\nu = 0, 1, 2, 3$ depending on atoimic number $Z$.}\label{fig2}
\end{figure}

It is possible to write down the spherically symmetric potential field, in which the atomic nucleus goes, as power series   $A[-2+\frac{x^2}{3}-\frac{x^3}{3}+\frac{x^4}{20}-\frac{x^5}{90}+...]$, where $A=Z^{**}e^2/a$, $x=2\frac{r}{a}Z^{**}$. This function differs from parabolic dependence and because of it non-harmonic amendments to the oscillations' energy arise. These amendments $(\Delta E_{\alpha\nu})$ for $\alpha$ - type of one-dimensional oscillations with oscillatory numbers $\nu = 0, 1, 2$ and $3$ are calculated in accordance with \cite{Flug1971}, vol. 2, p. 93, in the first and second orders of perturbation theory. As one would expect, the greatest values of amendments relate to the oscillatory condition with $\nu = 3$. Amendments to energy of inherent  $\alpha-$ type oscillations in conditions with  $\nu = 0, 1, 2$ and $3$ for various atoms are graphically submitted on Fig.~\ref{fig2}. 
On insertion of Fig.~\ref{fig2} these amendments are submitted in the other scale for atoms under $Z > 10$. 

\begin{figure}[ht]
\vspace*{0cm}
%\begin{center}
\includegraphics[width=8cm]{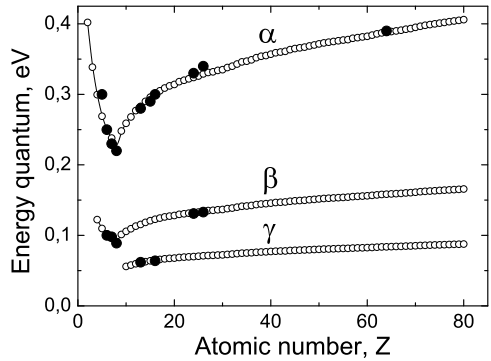}
%\end{center}
\vspace*{0cm}
\caption{Energy quantums of $\alpha-, \beta-, \gamma-$ oscillations depending on atomic number Z. Calculated values are shown by light circles. Experimental values are shown by dark circles.}
\label{fig3}
\end{figure}

The  $\beta-$ type of inherent oscillations, when the nucleus together with $K$  electrons participates in oscillations relatively other parts of the electron shell and the $\gamma-$ type of inherent oscillations when the nucleus together with $K$ and $L$  electrons participate in oscillations relatively other parts of the electron shell are also possible. Calculations have shown, that the energy spectrum of inherent harmonious oscillations  of  $\beta-$ and  $\gamma-$  types are described by the formula \eqref{eq7} with corresponding value of elementary quantum of oscillations~  $\hbar\omega_Z$ for each of these types that can be defined under the formula \eqref{eq9}, supposing $\Lambda = 0.2$~ for $\beta-$ type of inherent oscillations and $\Lambda=0.05$  for  $\gamma-$  type of inherent oscillations. The calculated and experimental values for energy quanta  $\alpha-, \beta- , \gamma-$ types of inherent oscillations depending on nuclear number $Z$, are submitted on Fig~\ref{fig3}.

Experiments show, that inherent oscillations of $\alpha-$, $\beta-$ and $\gamma-$ tipes are one-dimensional oscillations.  Consequently, for calculation of oscillation energy it is expedient to use the formula linear harmonious oscillator $E(\nu) = (1/2 + \nu)\hbar\omega_Z$, where $\nu = \nu_1 = 0,1,2...$ and $\nu_2 = \nu_3 = 0$ instead of eq. \eqref{eq7}.

It is established experimentaly, that "zero" energy of inherent oscillations $E (\nu=0) = (1/2)\hbar\omega_Z$, and also energies $(3/2)\hbar\omega_Z$ and $(5/2)\hbar\omega_Z$ participate in optical and electrical processes, that is typical of classical oscillations harmonious oscillators and forbidden for free quantum oscillators. It gives the basis to consider, that oscillatios of atomic nuclei in materials are not absolutely free and adiabatic, that meet conclusions of the quantum theory \cite{Born1927, Dirac1930, Stone1975, Dav1973, Dolt2002} about impossibility  to carry out strictly the adiabatic principle in materials. In this connection inherent nuclei oscillations in materials show dualism of physical properties, they manifest classical and quantum properties.

Inherent oscillations of nuclei cause strong electron-phonon interaction and stimulate corresponding features of physical properties of materials  \cite{Vd1999, Vd2000, Vd2002}. In adiabatic approach, as we see the nuclei oscillations are possible, but an electron-phonon interaction is excluded, though in reality this interaction accompanies inherent oscillations of nuclei. The given contradiction exists only in adiabatic electronics and absent in nonadiabatic electronics.

\section{Nonadiabatic electronics of materials} 

Nonadiabatic electronics differs in that it considers an exchange of energy between atomic nuclei and electrons to be an important feature of materials described by the operator $A$ in eq.~\eqref{eq2}. It is known that for the first time researches in non-adiabatic electronics have been undertaken more than 50 years ago by K. Huang \cite{Hua1950}, S. Pekar \cite{Pekar1951}, \cite{Pekar1952} and by other researchers who were studying the local color centers in dielectric crystals. In particular, it has been shown, that optical and thermal properties of these centers are caused by electron-vibrational transitions in which various oscillations of crystals can participate together with electrons.  In average, $S$ quanta of lattice elastic oscillations of one type participate in each such transition. According to the estimations, constant $S$ may reach 150, but experimental values $S \leq 22$. In zero-defects crystals, $S << 1$, that corresponds to the adiabatic principle, and $S > 1$ corresponds to nonadiabatic principle. In this connection for nonadiabatic electronics any materials with defects of structure such as the color centers, having the electron-vibrational nature, are important. The opportunity of electron-vibrational processes in semiconductors in the beginning caused doubts, but later the electron-vibrational centers (EVC) have been found in semiconductors too. It appeared that EVC and the electron-vibrational transitions in semiconductors, associated with EVC, are the cause of the characteristic phenomena such as phonon drag by electrons at Debye temperatures of phonons \cite{Vd2003}, thermal superconductivity, and also the hyperconductivity representing a version of superconductivity, that arises and exists at higher the transition temperatures than hyperconductivity, nearby the room temperatures and higher \cite{Vd2005}. These new properties of semiconductors, undoubtedly, relate to the nonadiabatic electronics, because they are caused by exchange of energy between electrons and atomic nuclei by means of electron-vibrational transitions between stationary vibrational states of nuclei \cite{Vd2006}.

It is well known, that thermoelectric power (TEP) or Zeebeck effect in a materials include contributions from electronic effects (the drift TEP) and electron-phonon effects ("phonon drag" TEP, or PDE): $V=V_d +V_{ph}$. The drift TEP ($V_d$) is caused by diffusion of electrons and holes under a gradient of temperature. The PDE ($V_{ph}$) exist due to dragging of electrons and holes by phonons stream. Thermoelectric power coefficient is sum of drift thermoelectric power coefficient and PDE coefficient: $\alpha = \alpha_d + \alpha_{ph}$,  according to \cite{Gur1946, Geb1953, Herr1953}. PDE was observed experimentally in Ge monocrystals only, at low temperatures \cite{Fr1953, Geb1953, Geb1954}. It forms a band of thermoelectric power and achieves a maximum between 15 K and 30 K. Calculations by C. Hrrring \cite{Herr1954} predicted the decrease of PDE, observable on experience at heating of material from 30 K up to 70 K, basically due to decreasing the  interaction between electrons and phonons. It has given the basis to wrongly believe, that PDE may exist only at low temperatures and to count completed its researches \cite{Czi1990}. Really, thermoelectric power of crystalline ropes of carbon nanotubes at temperatures from 4.2 K up to 300 K in laboratory of Nobel winner R. Smolli was presumably explained as PDE \cite{Hone1998}. It is possible due to strong electron-phonon interaction on EVC. Moreover, it was shown  \cite{Vd1999} that the PDE exist in semiconductors (containing EVC) as narrow bands of temperature-dependent thermoelectric power, located at Debye's temperatures of various phonons of a material.

It is possible to strengthen a connection of electrons with phonons in nonmetallic materials by introducing into them EVC. Strong interaction of electrons with phonons is provided due to inherent oscillations of nuclei in atoms of EVC. In such conditions, mobile electrons and holes are localised on EVC, the drift TEP decreases or disappears in general, and PDE dominates. The moving of electrical charges in a material under action of temperature gradient occurs basically as electron-vibrational transitions between EVC or between EVC groups, and the value $V_{ph}$ depends on speed of these transitions.

S. Pekar \cite{Pekar1953} described the intracenters nonradiative electron-vibrational transitions, caused by elastic oscillations of a crystal at frequency $\omega$. This theory is important for description of PDE as it is applicable for transitions between EVC, because in each material the EVC centers are similar each other and indiscernible. Speed of nonradiative transitions of center $\upsilon(\omega)$ from a condition k to a condition l reaches a maximum on frequency $\omega_m \cong S\omega_j$, where $\omega_j$ - frequency of phonon such as j, participating in transition, and it may be expressed by the following function:  
\begin{equation}
\label{eq10}
\upsilon(\omega)= \upsilon_m exp\left\{-\frac{(\omega-\omega_m)^2}{2g^{"}}+\frac{1}{6}\frac{g^{'''}(\omega-\omega_m)^3}{{g^"}^3})+...\right\},
\end{equation}
where $\upsilon_m$  - the maximal value $\upsilon(\omega)$, $g^" = \sum_j (q_{jk}-q_{jl})^2\omega_j^2 (n_j^* +1/2)$, $g^{'''}=\frac{1}{2}\sum_j(q_{jk}-q_{jl})^2\omega_j^3$, $(q_{jk}-q_{jl})$ - change of equilibrium coordinates of the center, $n_j^*$	  - average value of oscillatory quantum number. Near to a maximum of function eg. \eqref{eq10} value ($\omega - \omega_m$) is small, power row in a parameter exhibitors converges quickly and it is possible to be limited to the first square-law member of the row. 

Then,  according to the Debye rule, we having defined a temperature of a material $T=\hbar\omega/k$, a temperature of a material at the maximal speed of transitions $T_m = \hbar\omega_m /k$, and $\Theta=\hbar\sqrt{g^{"}}/k$. Believing, that the value of PDE is proportional to $\upsilon(\omega)$, it is possible to write down the temperature dependence of PDE as Gauss function 
\begin{equation}
\label{eq11}
\alpha_{ph}(T)= const \left\{-\frac{1}{2} \frac{(T-T_m)^2}{\Theta^2}\right\},
\end{equation}
where $\textit{const}$ is independent of temperature. We used this function eq.~\eqref{eq11} for approximation contours of experimental PDE bands in containing EVC materials, by selecting values $\textit{const}$ and $\theta$. 

\begin{figure}[ht]
\vspace*{0cm}
%\begin{center}
\includegraphics[width=8cm]{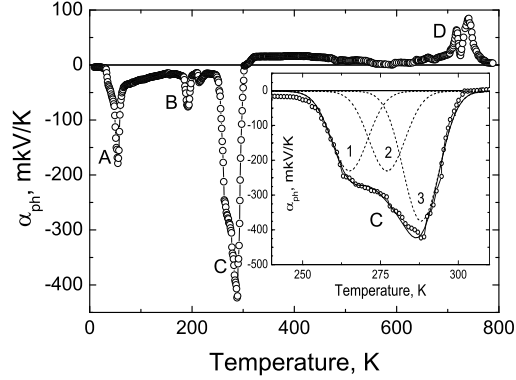}
%\end{center}
\vspace*{0cm}
\caption{Temperature dependency of thermoelectric power in Si, showing phonon drag effect as bands A, B, C, D. On insertion, the dashed curves 1, 2, 3 were calculated under eq.~\eqref{eq10}. Continuous curve is sum of the dashed curves. It approximate contour of band C.} 
\label{fig4}
\end{figure}

Typical temperature dependence of thermoelectric power in Si monocrystal with concentration EVC about $10^{15} sm^{-3}$, having rather narrow bands A, B, C, D with complex contours, is submitted on Fig.~\ref{fig4}. Dashed lines 1, 2, 3 on an insertion of Fig.~\ref{fig4} represent Gauss curves eq.~\eqref{eq10}, corresponding to longitudinal and transverse acoustic phonons in Si. Solid line is sum of the dashed lines. It describes a contour of a band C. It is noticed, that PDE bands exist only in materials containing EVC and in various materials they are located at Debye temperatures of acoustic and optical phonons of a material. It was observed in thin epitaxial layers of materials on substrates that the PDE bands located at Debye temperatures of a substrate phonons. These bands are caused by electron-vibrational transitions between EVC or between EVC groups and represent the PDE. 
%%%%
Components of PDE bands are well described by eq.~\eqref{eq10} in a vicinity of their extrema. Values $\Theta$  are identical to the contribution from any phonon type in each material and do not depend on temperature. Values $\Theta$ for some of materials are submitted in Tab.~\ref{tab1} and in Tab.~\ref{tab2}. They definitely reflect processes in the materials, independent from external conditions. Undoubtedly, here there are processes of an exchange by energy between systems of electrons and atomic nuclei which are characteristic for nonadiabatic electronics. 
\begin{table}[ht]
\caption{Values $\Theta$ in some monocrystal materials}\label{tab1}
~
\begin{center}
\begin{tabular}{|l|l|}
\hline
Material~~~~ $\Theta,~ K$ & Material~~~~ $\Theta,~ K$ \\[4pt]
\hline
GaAs~~~~~~~~~ 3.5        & InP~~~~~~~~~~~~ 3.6 \\
InAs~~~~~~~~~~ 4.5       & GaP~~~~~~~~~~~ 5.0 \\
InSb~~~~~~~~~~ 4.0       & graphite~~~~~ 5.0  \\
Ge~~~~~~~~~~~~~ 5.8      & CdHgTe~~~~ 14.3    \\ 
Si~~~~~~~~~~~~~~ 7.5     & ~~-~~~~~~~~~~~~~~~ - \\
\hline
\end{tabular}
\end{center}
\end{table}
\begin{table}[ht]
\caption{Values $\Theta$ in some single crystal films on substrates}\label{tab2}
~
\begin{center}
\begin{tabular}{|c|c|c|}
\hline
Material    & Substrate   &  $\Theta$,~ K  \\ [4pt]
\hline
InAs        & GaAs     &   3.5    \\

InSb        & GaAs     &   4.0    \\

Si          & sapphire &   4.7    \\

Carbon      & quartz   &   5.0    \\

nanotube    & fluorite &   5.0    \\

films       & YAl garnet     &   2.6    \\
\hline
\end{tabular}
\end{center}
\end{table}

Contrary to for a long time ratified opinion, the received results convince us that researches of PDE are not completed. The PDE researches are in the beginning of its developments, as well as nonadiabatic electronics as a whole.

\begin{figure}[ht]
\vspace*{0cm}
%\begin{center}
\includegraphics[width=8cm]{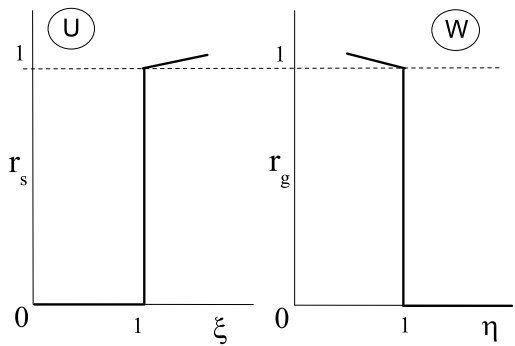}
%\end{center}
\vspace*{0cm}
\caption{Normalized temperature dependences of resistivity for a superconductor ($r_s$) and for a hyperconductor ($r_g$) on complex planes $U$ and $W$.}
\label{fig5}
\end{figure}

The hyperconductivity phenomenon differs from well-known superconductivity by details of physical mechanism. The normalized temperature dependences of resistivity of the superconductor; hyperconductors are shown on Fig.~\ref{fig5} in the dimensionless units $r_s = \rho/\rho_s$  and $r_g = \rho/\rho_g$, on complex planes $U$ and $W$.
Materials resistivity in the beginning of transition to the state with $\rho = 0$,  at temperatures $T_s$  and $T_g$, are denoted as $\rho_s$  and $\rho_g$. Normalized temperatures for semiconductor and superconductor are denoted $\xi = T/T_s$ and $\eta = T/T_g$, correspondingly, where T - Kelvin temperature.

Complex plane $W$ can be projected to the complex plane $U$, for example, with the help of transformation
\begin{equation}
\label{eq12}
U=(1-W_0)/(ReW-W_0)
\end{equation}
so that the point $\eta = 1$  was projected to the point $\xi = 1$. The real value $W_0$ may be chosen so that both temperature dependences of $\rho$  have coincided with each other with the greatest accuracy in plane $U$. The basic opportunity of such superposing of the temperature dependences $\rho$ proves the possibility of using of the phenomenological description of superconductivity as well as for description of hyperconductivity. Distinctions of these materials states with zero value $\rho$  consist only in details of physical mechanisms. 

\begin{figure}[ht]
\vspace*{0cm}
%\begin{center}
\includegraphics[width=8cm]{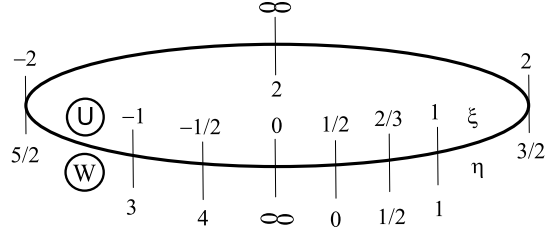}
%\end{center}
\vspace*{0cm}
\caption{Mutual conformity between temperatures are sitting in complex plane $U$ on axis $ \xi $ (for a superconductor) and in complex plane $W$ on axis $ \eta $ (for a hyperconductor).}\label{fig6}
\end{figure}

It is possible to consider axes $\xi$  and $\eta$ as lines leaving to positive and negative infinite large values, closed in infinity and forming the closed contours. Temperatures in terms of $\xi$ are laying above the axis on Fig.~\ref{fig6}. 
They relate to a superconductor. Temperatures in terms $\eta$ are laying below the axis of temperatures on Fig.~\ref{fig5}. They relate to a hyperconductor. Mutual conformity of the temperatures are sitting on the axes $\xi$  and  $\eta$, shown on Fig~\ref{fig6}. It is received with the help of transformation eq.~\eqref{eq12} under $W_0 = 2$. 

One can see from the Fig.~\ref{fig6} that temperature interval on axis $\xi$ in plane $Z$, where there is a superconductivity $(0 < \xi< 1 )$, corresponds to the interval $(-\infty < \eta < 1)$ on the axis $\eta$ lying in plane $U$, where there is a hyperconductivity and there are negative absolute temperatures. In other words, hyperconductors may be characterized by the negative absolute temperatures that lie higher than indefinitely high temperatures and may be applied to the description of physical systems with inverse population of energy levels as was shown in \cite{Lan1976}. In our case negative absolute temperatures should be related to the inverse population of EVC electron-vibrational states. 

Inherent oscillations of atomic nuclei influence on thermal, electric, optical and other physical properties of nonorganic and organic materials, fullerens, carbon nanotubes and carbon nanotube films. Inherent oscillations of nuclei, apparently, majorly define signals propagation in nerve fibres, action of some poisons, presence and properties of aura, and also various physical features of existence of live organisms and life in general. Thus, nonadiabatic electronics of materials promises to be less expensive, but not less various and useful, than existing adiabatic electronics.

\section{Discussion}

Fundamental opportunities of nuclei oscillations in atoms of materials are obvious from Schrodinger equation for a material and from adiabatic theory \cite{Born1927}. Nevertheless, these oscillations and the properties of materials related to them practically were not under investigation for decades, and implementations were limited by adiabatic electronics. Researches of the materials properties related to the transitions between electronic states \cite{Dolt2002} that are accompanied by changes of equilibrium positions or oscillations frequencies of atoms or ions in a material were related to the nonadiabatic electronics. Thus, displacement and oscillations of atoms or ions as the whole were considered, but one does not take into account possible oscillations of nucleus in atoms though they correspond in greater degree to the nonadiabatic principle. Modern nonadiabatic molecular mechanics, as a matter of fact, consider transitions between the basic and excited electron states, and movements of nuclei, described by classical equations of Newton. They do not take into the account the features of interaction between moving electric charges, loses nuclei quantum features and miss new properties of the materials caused by oscillations of atomic nuclei.

Aggregation of classical and quantum-mechanical principles in materials electronics testifies to inevitability of spreading of the classical mechanics into material microcosm what was predicted in the beginning of the last century by far-sighted scientists. 

Presently, the classical Newton mechanics is successfully applied to calculation of all electronic quantum states in atoms of any type \cite{Such2001}. It proves to be the most consent with experiment, uses the simple mathematical tool, and does not require difficult calculations. Apparently, soon problems related to molecules, fluids and solids will be solved by classical methods. Then electronics of materials will become classical, not split to adiabatic and nonadiabatic electronics, because such a division exists only in wave quantum mechanics of materials. 

\section{Conclusion} 

Systems of mutually bound particles and quasi-particles, including fermions and boson, consisting of electrons, holes, phonons and inherent oscillations of atomic nuclei, may exist in materials. Presence of the electron-vibration centers (EVC) in materials provokes formation of such fermion-boson systems, so far as, according to the theory and experiments, significant number of electrons, holes, phonons and inherent nuclei oscillations may have high concentration on such centers. Though atomic nuclei practically do not migrate within volume of a material, nevertheless, waves of inherent nuclei oscillations, as a result, cause the fermion-boson system to be mobile as a whole, giving to it properties of a fluid. Migration of such a fluid in a homogeneous material apparently doesn't cause energy consumption, so it may manifest property of superfluidity. Its movement may be considered as the mutually bound streams of fermions and bosons. The bosons stream provides thermal superconductivity, and the fermions stream provides hyperconductivity of a material. Therefore, effects of thermal superconductivity and hyperconductivity are linked together, as they accompany each other. Electron-vibrational transitions between various EVC create the phonon-drag effect on electrons at define Debye's temperatures of material phonons or substrate phonons.

\end{document}